\documentclass{ws-procs9x6}
\def\Journal#1#2#3#4{{#4}, {#1}, {#2}, #3} 
\newcommand{\etal}{et alii,\,}
\newcommand{\AMS}{\textsf{AMS}} 

\usepackage{verbatim}

\begin{document}

\title{AMS Observations of Light Cosmic Ray Isotopes \\and Implications for their Production in the Galaxy}
\author{Nicola Tomassetti}
\address{Perugia University and INFN, I-06122 Perugia, Italy\\
$^{*}$E-mail: nicola.tomassetti@pg.infn.it}

%%%%%%%%%%%%%%%%%%%%%
\begin{abstract}  %%%
%%%%%%%%%%%%%%%%%%%%%
Observations of light isotopes in cosmic rays provide information on their origin and propagation in the Galaxy. 
Using the data collected by \AMS-01 in the STS-91 space mission, we compare the measurements on 
$^{1}$H, $^{2}$H, $^{3}$He and $^{4}$He with calculations for interstellar propagation and solar modulation. 
These data are described well by a diffusive-reacceleration model with parameters that match the B/C ratio data. 
Close comparisons are made within the astrophysical constraints provided by the B/C data and within the nuclear 
uncertainties arising from the production cross sections. Astrophysical uncertainties are expected to be dramatically 
reduced by future data, but nuclear uncertainties may represent a serious limitation of the model predictions.
A diagnostic test for the reliability of the models is given by ratios such as 
$^{2}$H/$^{3}$He, $^{6}$Li/$^{7}$Li or $^{10}$B/$^{11}$B.
\end{abstract}

\bodymatter

%Observations of light isotopes in cosmic rays provide information 
%on their origin and propagation in the Galaxy. 
%Using the data collected by \AMS-01 in the STS-91 space mission, we report our final results 
%on the isotopic composition of hydrogen and helium between 200 MeV and 1.4 GeV per nucleon. 
%These measurements are in good agreement with the previous data and set new 
%standards of precision. We discuss the role of isotopic composition data in modeling 
%the cosmic ray production, acceleration and diffusive transport in the Galaxy.

%%%%%%%%%%%%%%%%%%%%%%%%%%%%%
\section{Introduction}    %%%
\label{Sec::Introduction} %%%
%%%%%%%%%%%%%%%%%%%%%%%%%%%%%

The rare secondary isotopes $^{2}$H, $^{3}$He and LiBeB are %under-abundant in the cosmic ray (CR) sources and are 
produced by collisions of primary cosmic rays (CRs) such as $^{1}$H, $^{4}$He or CNO with the interstellar matter (ISM). 
Secondary to primary ratios such as $^{2}$H/$^{4}$He, $^{3}$He/$^{4}$He or B/C 
give us information on the propagation of CRs through the ISM.
In many CR propagation studies the key parameters are inferred using the B/C ratio
and used to predict the secondary production for other rare species ($\bar{p}$, $\bar{d}$, ...) 
under the implicit assumption that all CRs experience the same propagation histories\cite{Strong2007,Putze2010,Trotta2011}.  
It is therefore important to test the CR propagation with nuclei of different mass-to-charge ratios.
In this work we report the new \AMS-01 observations for the $^{2}$H/$^{4}$He and $^{3}$He/$^{4}$He ratios
and compare them with propagation calculations. We study how these data are described by the models  
consistent with the B/C ratio within their astrophysical uncertainties (related to the CR 
transport parameters) and nuclear uncertainties (intrinsic of the $^{2}$H and $^{3}$He production rates).

%%%%%%%%%%%%%%%%%%%%%%%%%%%%%%%%%
\section{Observations}        %%%
\label{Sec::AMSObservations}  %%%
%%%%%%%%%%%%%%%%%%%%%%%%%%%%%%%%%

\AMS-01 operated on 1998 June in a 10-day space shuttle mission, STS-91, at an altitude of $\sim\,$380 km.
The spectrometer was composed of a permanent magnet, a silicon micro-strip tracker, 
time-of-flight scintillators, an aerogel \v{C}erenkov detector and anti-coincidence counters\cite{AMS01Report2002}.
Results on isotopic spectra have been recently published\cite{AMS01Isotopes2011}
including the ratios $^{2}$H/$^{4}$He, $^{3}$He/$^{4}$He, $^{6}$Li/$^{7}$Li, 
$^{7}$Be/($^{9}$Be+$^{10}$Be) and $^{10}$B/$^{11}$B in the range $0.2\--1.4$ GeV of kinetic energy per nucleon. 
\begin{figure}[!htb]
\centering
\psfig{file=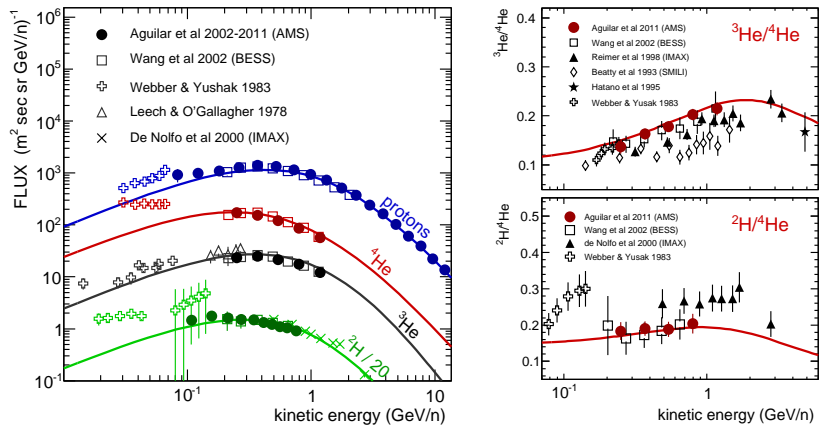,width=4.5in}
\caption{ 
  Left: energy spectra of CR proton, deuteron (divided by 20), $^{3}$He and $^{4}$He. 
  Right: isotopic ratios $^{2}$He/$^{4}$He and $^{3}$He/$^{3}$He.
  Calculations are shown together with the experimental 
  data\cite{AMS01Isotopes2011,DeNolfo2000,Reimer1998,Ahlen2000,Wang2002,Hatano1995,WebberYushak1983}.
  \label{Fig::ccIsotopeFluxes}
}
\end{figure}
Figure\,\ref{Fig::ccIsotopeFluxes} shows the \AMS-01 energy spectra of proton, deuteron, 
helium isotopes, and the ratios $^{2}$He/$^{4}$He and $^{3}$He/$^{3}$He. These data are free from atmospheric background.
Other data come from balloon borne experiments\cite{DeNolfo2000,Reimer1998,Ahlen2000,Wang2002,Hatano1995,WebberYushak1983}.  

%%%%%%%%%%%%%%%%%%%%%%%%%%%%%%%%%%%%%%%%%%%%%%%%%%%%
\section{Cosmic Ray Transport and Interactions}  %%%
\label{Sec::CRPropagation}                       %%%
%%%%%%%%%%%%%%%%%%%%%%%%%%%%%%%%%%%%%%%%%%%%%%%%%%%%

The Galactic CR transport is characterized by diffusion in the turbulent magnetic field,
nuclear interactions, decays, energy losses and diffusive re-acceleration.
We describe the data using
\texttt{GALPROP-v50.1} which numerically solves the CR propagation equation in a 
cylindrical diffusive region for given source and matter distributions\cite{Trotta2011}.
We adopt the ``conventional model'' which finely reproduces the primary CR fluxes and the B/C ratio at intermediate energies
under a diffusion-reacceleration scenario.
We model the heliospheric propagation under the \textit{force-field} approximation\cite{Gleeson1968},
using the parameter $\phi=500$\,MV to characterize the modulation strength for 1998 June.
To compute the $^{2}$H and $^{3}$He production rate in the ISM, we consider the reactions 
$^{4}$He$\rightarrow$$^{3}$He, $^{4}$He$\rightarrow$$^{3}$H, $^{4}$He$\rightarrow$$^{2}$H,
CNO$\rightarrow$$^{3}$He, and $p+p \rightarrow \pi + ^{2}$H (the $^{3}$H isotopes subsequently decay into $^{3}$He).
Most of the reactions involve the $^{4}$He spallation cross sections (CSs) 
that we have adapted from the parametrizations of \cite{Cucinotta1993}.
Contributions from of heavier nuclei 
(CNO or Fe, giving $\sim$\,5\% of the flux) are included using the CS data from
the latest \texttt{GALPROP-v54} version\cite{Vladimirov2011}. 
For collisions with He targets ($\sim$\,10\% of the ISM) the algorithm of \cite{Ferrando1988} is used.
\begin{figure*}[!htb]
\centering
\psfig{file=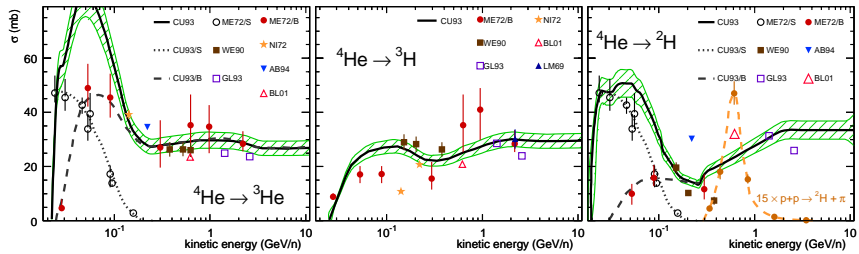,width=4.5in}
\caption{ 
  CS parametrization\cite{Cucinotta1993} for the channels $^{4}$He$\rightarrow$$^{3}$He,
  $^{4}$He$\rightarrow$$^{3}$H, $^{4}$He$\rightarrow$$^{2}$H
  and $p+p \rightarrow \pi + ^{2}$H (multiplied by 15). 
  The experimental data are found in \cite{Cucinotta1993,Meyer1972,Blinov2008}.
  \label{Fig::ccHeliumDeuteronXS}
}
\end{figure*}
We assume the \textit{straight-ahead} approximation to link the fragment-progenitor energies $E$ and $E'$
through $\sigma(E,E')\approx\sigma(E)\delta(E-E')$,
that is valid at some percent of accuracy when the progenitor is heavier than the fragment\cite{Kneller2003,Cucinotta1993}.
For the $p$-$p$ fusion channel the kinetic energy per nucleon is not conserved:
we assume $\sigma(E,E')\approx\sigma(E)\delta(E-\xi E')$, where $\xi\approx$\,4 is the average 
inelasticity for the $^{2}$H production\cite{Meyer1972}.
This reaction contributes to the $^{2}$H flux at $E \lesssim$\,250\,MeV\,nucleon$^{-1}$. 
The main CSs are shown in Fig.~\ref{Fig::ccHeliumDeuteronXS} together with the data.  
For $^{2}$H and $^{3}$He the contributions of break-up (B) and stripping (S) reactions are shown separately.
Predictions for CR spectra and for the ratios $^{2}$H/$^{4}$He and $^{3}$He/$^{4}$He under this setup 
are reported in Fig.\,\ref{Fig::ccIsotopeFluxes}.

%%%%%%%%%%%%%%%%%%%%%%%%%%%%%%%%%%%%%%%%%%%%%%%%%%%%%%%%%%%%%%%%%%%%%%%%%
\section{Model Uncertainties for the $^{2}$H and $^{3}$He Productions} %%%
\label{Sec::ModelUncertainties}                                       %%%
%%%%%%%%%%%%%%%%%%%%%%%%%%%%%%%%%%%%%%%%%%%%%%%%%%%%%%%%%%%%%%%%%%%%%%%%%

\begin{figure}[!htb]
\centering
\psfig{file=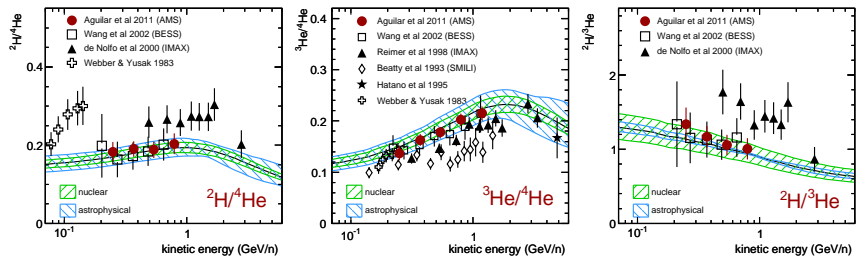,width=4.5in}
\caption{ 
  Astrophysical and nuclear uncertainty bands for the predicted ratios $^{2}$H/$^{4}$He, $^{3}$He/$^{4}$He and $^{2}$H/$^{3}$He
  in comparison with the experimental data\cite{AMS01Isotopes2011,DeNolfo2000,Reimer1998,Ahlen2000,Wang2002,Hatano1995,WebberYushak1983}.
  \label{Fig::ccIsotopeRatios2X3}
}
\end{figure}

We consider two classes of uncertainties. 
The \textit{astrophysical uncertainties} are related to the transport parameters constrained by the B/C data. 
The relevant parameters are $\delta$ (diffusion coefficient spectral index), 
$v_{A}$ (Alfv\'enic speed) and the ratio between $D_{0}$ (diffusion coefficient normalization) and $L$ (halo height). 
We have performed a grid scan in the parameter space $\{\delta, v_{A}, D_{0}/L\}$ by running \texttt{GALPROP}
several times, while the other inputs (\textit{e.g.} source parameters or $\phi$) are kept fixed. 
In order to derive the astrophysical errors for the $Z\leq 2$ predictions,
we select the models compatible with the B/C data within one sigma of uncertainty.
Our purpose is estimating the parameter uncertainties rather than determining their 
values (\textit{e.g.}, as recently done in \cite{Coste2011}).
The \textit{nuclear uncertainties} on the $^{2}$H and $^{3}$He calculations 
are those arising from uncertainties in their production CSs.
In order to estimate these uncertainties using the CS data,
we re-fit the normalization factors of their parametrizations.
The error bands are shown in Fig.\,\ref{Fig::ccHeliumDeuteronXS} for the main 
production CSs of $^{2}$H and $^{3}$He. Their corresponding errors in the predicted ratios 
$^{2}$H/$^{4}$He and $^{3}$He/$^{4}$He are shown in Fig.\,\ref{Fig::ccIsotopeRatios2X3} in
comparison with the \textit{astrophysical uncertainty} bands.
The \AMS-01 data agree well with the calculations within the \textit{astrophysical uncertainties}, 
indicating a good consistency with the B/C-based propagation picture.
It is also clear that these ratios carry valuable information on the CR transport parameters
and can be used to tighten the constraints given by the B/C ratio.
On the other hand, the \textit{nuclear uncertainties} represent an intrinsic limitation on the accuracy of the predictions.
Unaccounted errors or biases in the CS estimates cause errors on the predicted ratios which, 
in turn, may lead to a mis-determination of the CR transport parameters. 
For high precision data upcoming from PAMELA or \AMS-02, CS errors may become the dominant source of uncertainty.  
A strategy to test the model consistency with CR data is given by the comparison with 
secondary to secondary ratios such as $^{2}$H/$^{3}$He.
In fact the $^{2}$H and $^{3}$He isotopes have similar astrophysical origin,
so that their ratio is almost insensitive to the propagation physics and 
can be used to probe the net effect of nuclear interactions. 
Thus, a mis-consistency between calculations and $^{2}$H/$^{3}$He data would indicate 
systematic biases in the CSs that cannot be re-absorbed by a different choice of the propagation parameters. 
From Fig.\,\ref{Fig::ccIsotopeRatios2X3}, the nuclear uncertainty in the $^{2}$H/$^{3}$He ratio is larger than the astrophysical one.  

%%%%%%%%%%%%%%%%%%%%%%%%%%%%%%
\section{Conclusions}      %%%
\label{Sec::Conclusions}   %%%
%%%%%%%%%%%%%%%%%%%%%%%%%%%%%%

We have compared the \AMS-01 observations of the $^{2}$H/$^{4}$He and $^{3}$He/$^{4}$He ratios in CRs 
with model predictions for their production in the ISM. 
These ratios are well described by propagation models consistent with the B/C ratio,
suggesting that He and CNO nuclei experience similar propagation histories.
The accuracy of the secondary CR calculations depends on the reliability of the CSs employed.
CS parametrizations may be improved using more refined calculations or more precise accelerator data.
The use of ratios such as $^{2}$H/$^{3}$He, $^{6}$Li/$^{7}$Li or $^{10}$B/$^{11}$B can represent
a possible diagnostic test for the reliability of the calculations: any CR propagation model,
once tuned on secondary to primary ratios, must correctly reproduce the secondary to secondary ratios as well.
Precision modeling may also require a more refined solar modulation description in place of the 
simple \textit{force-field}. This aspect can be better inspected by the \AMS-02 log-term observations.

%%%%%%%%%%%%%%%%%%%%%%%%%%%%%%%

\bibliographystyle{ws-procs9x6}
\bibliography{ws-pro-sample}

\end{document}